\documentclass[twocolumn,noshowpacs,preprintnumbers,amsmath,amssymb]{revtex4}

\usepackage{verbatim}

\usepackage{graphicx}
\usepackage{dcolumn}
\usepackage{bm}
\usepackage{wrapfig}

\begin{document}

\title{Synthesis and thermoelectric characterization of Bi2Te3 nanoparticles}

\author{Marcus Scheele}
\email{scheele@chemie.uni-hamburg.de}
\author{Andreas Kornowski}
\author{Christian Klinke}
\author{Horst Weller}
\affiliation{Institute of Physical Chemistry, University of Hamburg, Germany}
\author{Niels Oeschler}
\author{Katrin Meier}
\affiliation{Max-Planck-Institute for Chemical Physics of Solids, Dresden, Germany}

\begin{abstract} % abstract goes here

We report a novel synthesis for near monodisperse, sub-10-nm Bi2Te3 nanoparticles. At first, a new reduction route to bismuth nanoparticles is described which are applied as starting materials in the formation of rhombohedral Bi2Te3 nanoparticles. After ligand removal by a novel hydrazine hydrate etching procedure, the nanoparticle powder is spark plasma sintered to a pellet with preserved crystal grain sizes. Unlike previous works on the properties of Bi2Te3 nanoparticles, the full thermoelectric characterization of such sintered pellets shows a highly reduced thermal conductivity and the same electric conductivity as bulk n-type Bi2Te3.

\end{abstract}

\maketitle

\section{Introduction}

Bulk Bi2Te3 and solid solutions thereof are the key materials for state-of-the-art thermoelectric (TE) devices at room temperature. The efficiency of such devices is defined as

\begin{equation}
zT = \frac{\sigma S^{2} T}{\kappa _{L} + \kappa _{e}}
\end{equation}

and peaks at 1.14 for bulk (Bi2Te3)0.25(Sb2Te3)0.72(Sb2Se3)0.03~\cite{1}. It is estimated that a three-fold increase would result in a Carnot efficiency similar to those of conventional heat generators, thus making thermoelectric materials a promising subject in the search for new power generators. 

With the electric conductivity $\sigma$, the Seebeck coefficient (or thermopower) $S$, the absolute temperature $T$, and the lattice and electronic part of the thermal conductivity $\kappa_{L}$ and $\kappa_{e}$, zT can only be significantly changed by varying the Seebeck coefficient or the lattice thermal conductivity. As predicted theoretically~\cite{2,3} and demonstrated experimentally~\cite{4}, both parameters can be manipulated by nanotechnology. 

Specifically, nanoparticles have been predicted to show a strong scattering effect on phonons similar to that of atomic impurities or crystal boundaries~\cite{5}. The effect was found to be inversely related to the nanoparticle diameter. An instructive summary on the effect of nanoscalic dimensions on thermoelectric materials has been given by Dresselhaus et al.~\cite{6}.

Consequently, several recent reports have demonstrated experimental evidence on the perspectives of nanostructured materials for thermoelectric applications, taking advantage of the phonon scattering in these materials~\cite{7,8}.

Particularly important for large-scale applications was a work on hot-pressed bismuth antimony telluride nanoparticles fabricated by ball-milling, which yielded zT of 1.4 at 400 K~\cite{9}. In a similar approach, solution grown bismuth telluride nanostructures of approximately 30 nm in diameter were included into a Bi2Te2.7Se0.3 bulk matrix in various concentrations, which led to a decrease in lattice thermal conductivity (1.2 WK$^{-1}$m$^{-1}$) meanwhile maintaining low resistivity (3.7 m$\Omega$cm)~\cite{10}. Both of these works applied hot-pressing of nanoparticles which is a promising approach towards nanostructured bulk materials provided the grain sizes do not increase greatly during the treatment. Particularly suitable in this respect is sintering with a pulsed direct current, known as spark plasma sintering (SPS). This method allows for relatively short sintering times which minimizes the post-synthetic crystal grain growth of nanoparticles~\cite{11}.

Since phonon scattering depends on the size and shape of the nanoparticles, a good control over these parameters is essential in achieving further improvements in thermoelectric efficiencies. Several solution-based attempts to a more controlled synthesis of large amounts of small, crystalline bismuth telluride nanoparticles have been reported. For example, crystalline and uniform bismuth telluride disks of 100-200 nm in diameter~\cite{12}, 2.5 to 10 nm bismuth nanoparticles of medium crystalline quality~\cite{13} and highly crystalline bismuth telluride/bismuth sulphide core/shell nanorods of 35 to 290 nm in lengths~\cite{14} have been reported. A recent work presented a breakthrough in the size-controlled synthesis of crystalline Bi2Te3 nanoparticles in the range of 17 nm to 100 nm of narrow size distribution~\cite{15}. Where the thermal conductivity of these nanoparticles pressed to a dense pellet could be reduced to 0.5~WK$^{-1}$m$^{-1}$ at room temperature, zT was only 0.03 due to a poor electric conductivity stemming from organic residues of the former stabilizing ligands.

\begin{figure}[htbp]
  \centering
  \includegraphics[width=0.45\textwidth]{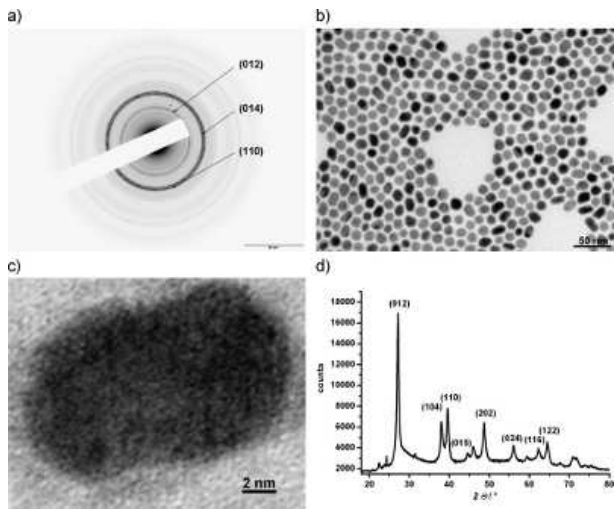}
  \caption{\textit{a) Selective-area electron diffraction (SAED) of bismuth nanoparticles; colors inverted for clarity. b) Transmission electron microscopy (TEM) and c) high-resolution TEM (HRTEM) of a single bismuth nanoparticle showing the (110) lattice spacing. d) Powder (p)-XRD indexing according to ICDD PDF-card no. 44-1246 [trigonal (R$\overline{3}$m)(166)].}}
\end{figure}

To date, a synthesis yielding amounts in the gram scale of uniform, sub-10-nm bismuth telluride nanoparticles of high crystallinity has not been reported.

In the first part of this paper, we report a synthesis which fulfils these requirements. In the second part we demonstrate the fabrication of a nanostructured bulk material comprising of the previously prepared particles by spark plasma sintering after carefully removing the particles' ligands by a novel hydrazine hydrate based etching procedure. In contrast to other reported ligand removal techniques, the electric conductivity of thus purified nanoparticles is identical to the room temperature bulk value, which sets the basis for a large zT.

In the last part we will present our data on the characterization of the TE properties of such a material.

\section{Results and discussion}

The difficulty in solution-processed bismuth telluride nanoparticle synthesis is the high reactivity of tellurium with bismuth salts. Where Fang et al.~\cite{12} and Ramanath et al.~\cite{13} have nonetheless demonstrated impressive control over this reaction for large particle sizes, the high reactivity usually does not allow for a controlled growth of small bismuth telluride nanoparticles. An exception is the work by Badding et al.~\cite{15}, although even their synthetic procedure did not yield sub-10 nm particles.

The innovation of our synthesis is its two-step nature via a bismuth nanoparticle intermediate. We have tested several attempts for the synthesis of such an intermediate (see supporting information) and found it best to apply oleylamine as the reducing agent at 60$^{\circ}$C together with 1-docdecanethiol (DDT) as the stabilizer (Figure 1a-d).

\begin{figure}[htbp]
  \centering
  \includegraphics[width=0.45\textwidth]{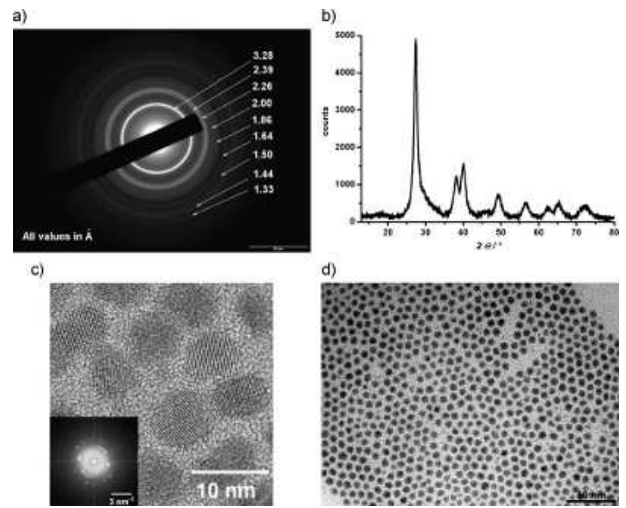}
  \caption{\textit{Typical bismuth tellurium alloy nanoparticles; (a) SAED with measured d-values; (b) p-XRD; (c) HR-TEM with corresponding FFT; the predominant lattice spacing is 3.28 $\pm$ 0.03 A; (d) TEM.}}
\end{figure}

The as prepared bismuth nanoparticles can be treated with a solution of tellurium in trioctylphosphine (TOP-Te) without further purification in the same batch to yield a bismuth-tellurium alloy (Figure 2a-d). These alloy nanoparticles are single-crystalline and possess a mean particle diameter only $\sim$20~\% larger than the former bismuth species due to the inclusion of tellurium (Figure 3a-d).

\begin{figure}[htbp]
  \centering
  \includegraphics[width=0.45\textwidth]{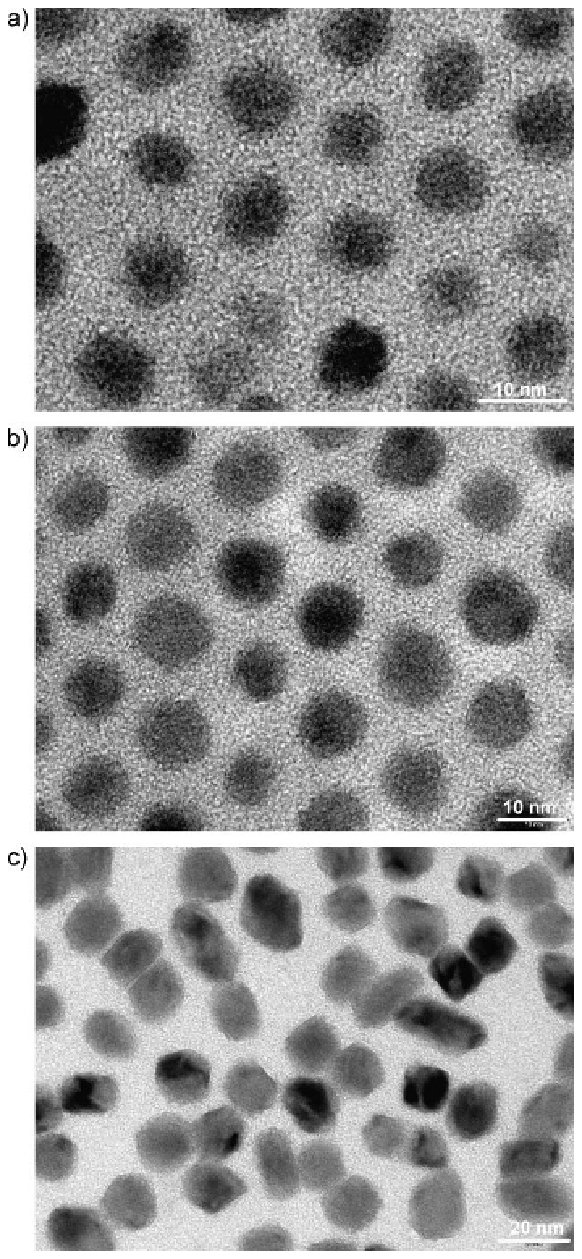}
  \caption{\textit{Bismuth-tellurium nanoparticles (a) by a 30.0 mM solution with respect to the Bi precursor; (b) by a 90.0 mM solution with respect to the Bi precursor; (c) by a 30 mM solution and dilution with toluene.}}
\end{figure}

The randomness of this inclusion causes the broadening of the reflections in XRD measurements (Figure 2b and 4). Bos et al. have given an extensive overview of bismuth tellurides of various compositions and have assigned them to an infinitely adaptive series of the general formular (Bi2)m(Bi2Te3)n with (m+n) being a multiple of 3~\cite{16}. Their presented XRDs of the members of this homologous series show the very little deviations in XRD reflections and intensity as one moves along from pure bismuth to bismuth telluride in the Bi2Te3 modification. It is believed that the large flexibility of bismuth tellurides in terms of their chemical composition leads to the broad XRD reflections.

\begin{figure}[htbp]
  \centering
  \includegraphics[width=0.45\textwidth]{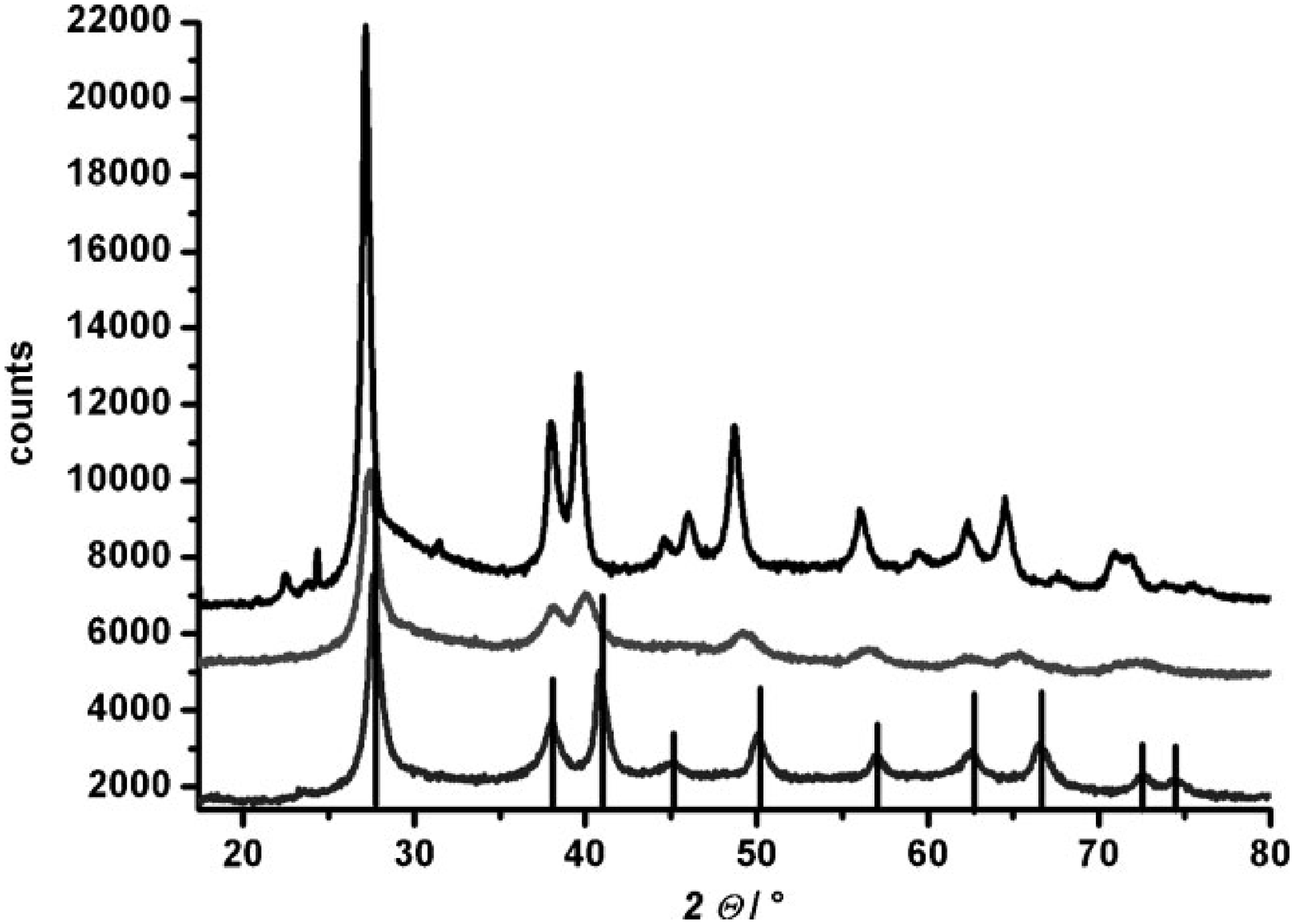}
  \caption{\textit{p-XRD patterns of (top) bismuth nanoparticles; (middle) the same nanoparticles after alloying with tellurium; (bottom) the alloy after annealing at 110$^{\circ}$C in solution; the black bars represent characteristic reflections for rhombohedral Bi2Te3 [(R$\overline{3}$m)(166), PDF-Card 15-0863].}}
\end{figure}

Under the specified conditions, the particles can be kept for weeks without any changes in size, composition and crystalline phase. When such a solution is heated to 110$^{\circ}$C for 18 hours, the particle size does not undergo significant changes. However, the shape changes slightly from almost spherical to a more rhombohedral structure and the XRD reflections shift towards practically single-phase, rhombohedral Bi2Te3 (space group R$\overline{3}$m 166) provided an appropriate amount of tellurium was added (Figure 4 and 5a-d). (The sharp reflections unknown to Bi2Te3 are assigned to excess ligand, which is explained in more detail in the supporting information.) This way, we could synthesize Bi2Te3 nanoparticles in a range of 7 to 50 nm in diameter depending on the size of the bismuth nanoparticles used as the starting material.

\begin{figure}[htbp]
  \centering
  \includegraphics[width=0.45\textwidth]{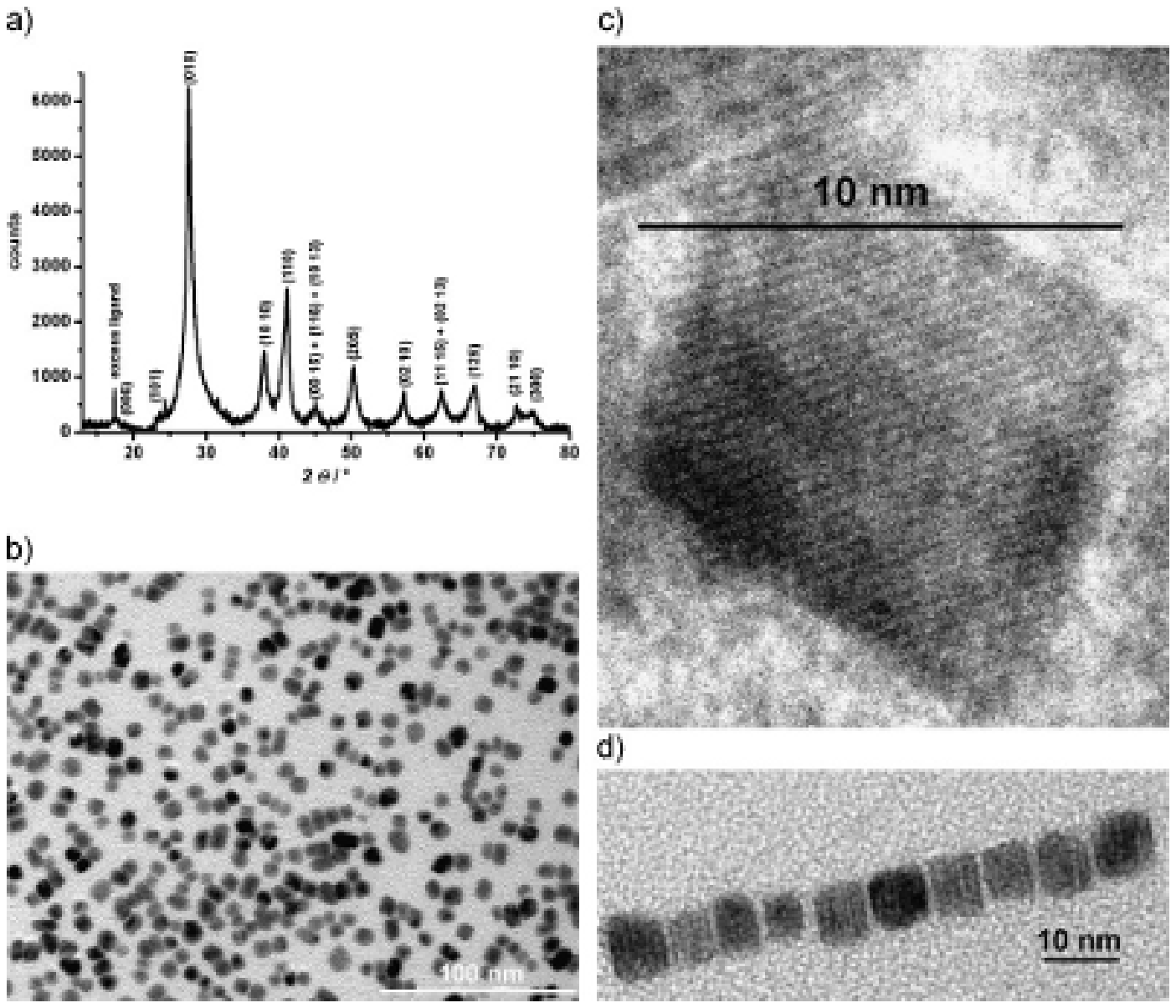}
  \caption{\textit{Bi2Te3 nanoparticles (a) p-XRD; indexing according to ICDD PDF-card No. 15-0863 (trigonal [R$\overline{3}$m)(166)]; (b) TEM; (c) + (d) HR-TEM of typical nanoparticles.}}
\end{figure}

To fabricate sintered pellets from these nanoparticles, we removed the nanoparticles' ligand shell via a two-step procedure. In the first step, the ligand shell of DDT was exchanged by a ligand shell of oleic acid. This exchange is favoured by the weak interaction of DDT with the nanoparticles' surface (see supporting information, Figure S5). In the second step, oleic acid stabilized Bi2Te3 nanoparticles were suspended in a hydrazine hydrate/hexane two-phase mixture. We chose hydrazine hydrate at this point for three reasons: i) As a base it readily deprotonates the acid leaving it incapable of binding to the nanoparticles' surface. ii) As a reducing agent it repairs oxide defects on the nanoparticles surface. iii) In comparison with the highly explosive and carcinogenic anhydrous hydrazine, hydrazine hydrate can be handled relatively safely in a fume hood under ambient conditions. 

As a result of the hydrazine treatment, the ligand-free, polar nanoparticles transferred into the polar phase, whereas the aliphatic hydrazinium oleate remained in the organic phase. Detailed NMR-studies of a similar process can be found elsewhere~\cite{17}. Drying of thus purified nanoparticles yielded a Bi2Te3 nanopowder. 

This nanopowder was spark plasma sintered to a macroscopic pellet. Provided a careful control of the sintering parameters, we observed the size and phase of the nanoparticles in the pellet to be the same as those obtained directly after synthesis. This is supported by HR-SEM imaging displaying small crystalline grains of $\sim$15 nm in a sintered sample of originally 14 nm big nanoparticles and SAED measurements where the fringe patterns indicate polycrystallinity (Figure 6). (For a more detailed study of the effect of different sintering parameters see supporting information.)

\begin{figure}[htbp]
  \centering
  \includegraphics[width=0.45\textwidth]{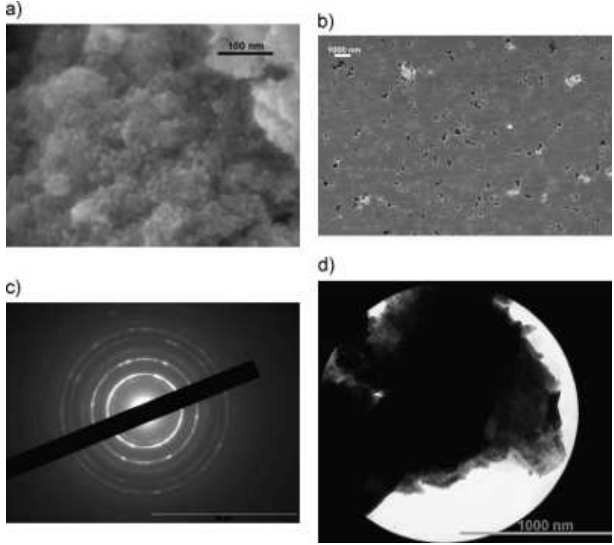}
  \caption{\textit{(a) SEM image of sintered Bi2Te3 nanoparticles (14 nm). (b) Low-magnification SEM image of the same sample displaying occasional voids. (c) SAED of a small piece of a nanoparticle pellet (d) TEM image of the piece.}}
\end{figure}

We now turn to the discussion of thermoelectric properties of the Bi2Te3 nanoparticles in the order i) electric conductivity, ii) thermopower and iii) thermal conductivity: 

Our transport measurements under dc conditions show a classic ohmic and semiconducting behaviour (Figure 7a). This is confirmed by ac resistivity measurements between 5 K and 300 K showing a 20~\% drop in resistivity as one moves along in this temperature regime (Figure 7b). With respect to what was reported about n-type bulk Bi2Te3~\cite{16} this is unusual as one should expect metallic behaviour. However, it has to be stressed here that the present material possesses a highly polycrystalline and granular structure. Such materials are known to show significantly altered transport properties as compared to their homogeneously disordered equivalents~\cite{18}. Still, the room temperature resistivity is less than 50~\% higher than that of n-type bulk Bi2Te3 ($\sim$1.4 m$\Omega$cm), which is unprecedented by solution-grown Bi2Te3 nanoparticles.

\begin{figure}[htbp]
  \centering
  \includegraphics[width=0.45\textwidth]{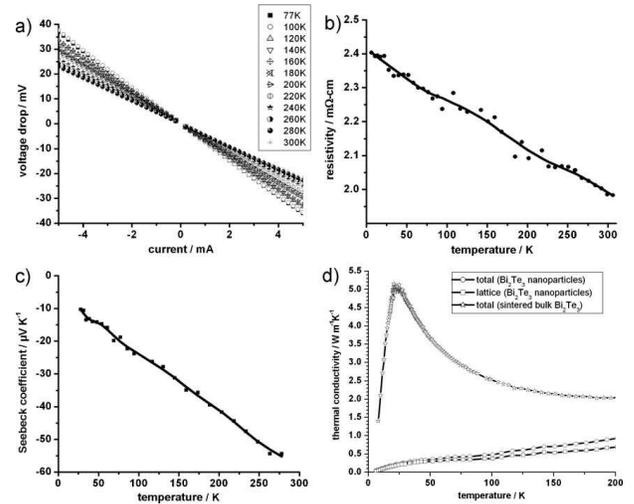}
  \caption{\textit{Transport properties of Bi2Te3 nanoparticles: (a) dc I/V-measurements (b) ac resistivity (c) thermopower (d) total and lattice thermal conductivity of a pellet of sintered nanoparticles with conditions "c" (Figure S6ii, Supporting Information). The thermal conductivity data includes a sintered bulk Bi2Te3 sample for comparison.}}
\end{figure}

The thermopower of the sintered Bi2Te3 nanoparticle pellet shows clear n-type behaviour (Figure 7c) and a room temperature absolute value of 60 $\mu$V$K^{-1}$ which compares to 180 $\mu$V$K^{-1}$ for n-type bulk Bi2Te3. This is fully in accordance with the work of Glatz and Beloborodov who predicted a decrease in thermopower "most effective for small grains" due to "the delicate competition of the corrections of thermoelectric coefficient and the electric conductivity"~\cite{18}. 

For reasons stated above, materials of small crystal grain sizes are most promising for thermoelectrics in terms of their thermal conductivity. This is immediately apparent from Figure 7d which shows the thermal conductivity of such a material. Due to the significant radiative heat loss above 200 K only the data below this temperature can be considered. We also display the lattice thermal conductivity on subtracting the electronic contribution by assuming the Sommerfeld value for the Lorenz number. For comparison we included the thermal conductivity of a sintered sample of commercially available bulk Bi2Te3. 

Due to the break-down of Umklapp-processes at low temperatures, single crystals of sufficiently large grain sizes show a maximum in thermal conductivity between 4 K and 9 K before dropping sharply when approaching the zero-point in absolute temperature~\cite{19}. Our homogeneously disordered bulk sample displays exactly this behaviour although the maximum occurs at a higher temperature and possesses a lower value due to the non-single crystalline nature. In contrast, the sintered Bi2Te3 nanoparticle pellets show a decrease in thermal conductivity without passing a maximum when falling below a certain temperature - in our case approximately 30 K. It is believed that at this temperature, the phonon mean free path approaches its maximum value and is restricted to it for all temperatures below the threshold temperature due to limited grain boundaries. If we apply the kinetic gas theory derived approximation for the thermal conductivity

\begin{equation}
\kappa_{L} = \frac{1}{3} c_{V} v l_{t}
\end{equation}

(where $\kappa_{L}$ is the lattice thermal conductivity, $c_{V}$ the specific heat for unit volume, $v$  the speed of sound and $l_{t}$ the phonon mean free path), it is apparent that for a constant $l_{t}$, the temperature dependence of the lattice thermal conductivity should be that of the specific heat. On using Debye's theory, for low temperatures the specific heat should have a T$^{2}$ to T$^{3}$ dependence, which is reasonably resembled by our measurements between 5 K and 30 K. Most importantly in the thermal conductivity comparison of the bulk to nanoparticle samples, we observe a minimum depletion by a factor 2 which increases to as much as one order of magnitude. This is in agreement with what was reported by Badding et al.~\cite{15} for a similar sample showing the validity of these measurements. 

In the last part of this communication, we discuss the effect of optimized fabrication parameters on the thermoelectric properties. In Figure 8 we prolonged the sintering duration in order to improve the thermopower of otherwise identical sintered Bi2Te3 nanoparticle pellets. We also present the measurements of a sintered Bi2Te3 bulk sample for comparison. Longer sintering times repeatedly yielded a higher absolute thermopower and a lower resistivity. The room temperature thermopower increased to 80 $\mu$VK$^{-1}$ and the resistivity decreased to 1.3 m$\Omega$cm which is identical with the value for n-type bulk Bi2Te3. Moreover, the temperature dependence of charge carrier transport is now metallic in accordance with the bulk material.
From SEM imaging it is apparent that the granular structure is unaffected by the longer sintering duration (Figure 8i and ii). As a result, we measure a lattice thermal conductivity of 0.8 Wm$^{-1}$K$^{-1}$ at 200 K, which is consistent with the measurement for short sintering durations in Figure 7d.

\begin{figure}[htbp]
  \centering
  \includegraphics[width=0.45\textwidth]{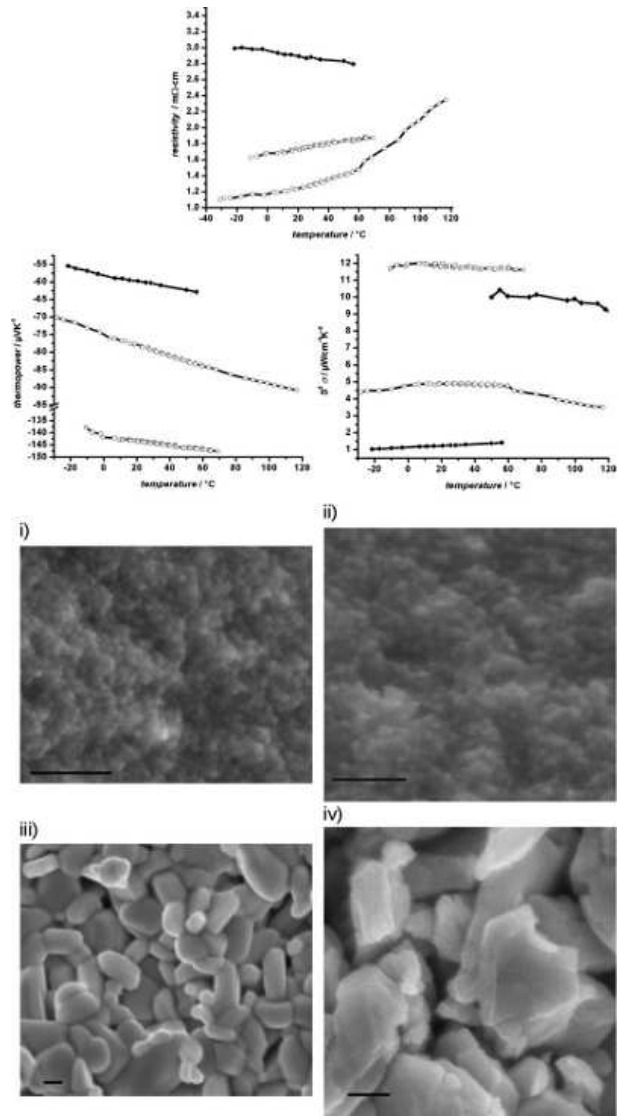}
  \caption{\textit{Resistivity, thermopower, power factor and morphology of sintered 9 nm Bi2Te3 nanoparticles.  Closed circles: Short sintering (conditions "c"), morphology "i"; Open circles: Long sintering (conditions "d"), morphology "ii"; Closed diamonds: after thermal treatment, morphology "iii"; open squares: Homemade bulk Bi2Te3 standard (SPS conditions "e"), morphology "iv". All scale bars equal 100 nm.}}
\end{figure}

For better comparison, we summarize the thermoelectric data so far acquired in this work together with suitable reference compounds reported in literature in Table 1.

When applying a post-sintering thermal treatment at 250$^{\circ}$C in a 0.1 bar helium atmosphere, the power factor of a typical  pellet of nanoparticles increases from 1 to almost 10 $\mu$Wcm$^{-1}$K$^{-2}$ (Figure 8). It is reported that at this temperature no melting is to be expected for Bi2Te3 nanoparticles~\cite{15,20}. Where SEM imaging suggests a significant reorganization of the nanoparticle pellet's structure (Figure 8iii), XRD measurements (see supporting information) reveal that the crystalline domains only grow slightly (~30 nm) during this process. It is therefore likely that each of the ~300 nm grains in Figure 8iii consists of multiple crystalline domains to maintain a low thermal conductivity. This result suggests that careful thermal treatment of the sintered Bi2Te3 nanoparticle pellets reported in this work has the potential to yield a zT of at least 0.4 mainly due to an increase in thermopower by optimized crystal grain sizes. This will be subject to a separated publication.

\begin{table*}[tbp]
\begin{center}
\caption {\textit{Comparison of thermoelectric parameters of selected Bi2Te3 samples. Note that all values were recorded at 300 K except for the thermal conductivity which can only be considered up to 200 K due to radiation effects at higher temperatures.}}
\vspace{0.5cm}
\begin{tabular}{|c|c|c|c|c|} \hline
Type of & Resistivity &	Thermopower &	Thermal conductivity & zT  \\ 
Bi2Te3 sample	& [m$\Omega$cm] (300 K) &	[$\mu$VK$^{-1}$] (300 K) & [Wm$^{-1}$K$^{-1}$] (200 K) & (300 K) \\ \hline
n-type bulk &	1.4~\cite{16} & 180~\cite{16} & 2.2 (sintered sample) & 0.32 \\ \hline
this work	& 1.3 & 80 & 0.8 & 0.2 \\ \hline
\cite{15} & 20 & 91 & 0.5 & 0.03 \\ \hline

\end{tabular}
\end{center}
\end{table*}

\section{Conclusions}

Because of the large reactivity of bismuth and tellurium, the key to unprecedented sub-10 nm single-crystalline Bi2Te3 nanoparticles of narrow size distribution is a bismuth nanoparticle intermediate as part of a one-pot two-step solution based procedure. To meet the key requirement for a sufficient electrical conductivity and thus a large power factor we introduced a novel hydrazine hydrate based ligand removal prior to the sintering of the nanoparticles to macroscopic pellets. This yielded an electrical conductivity which is virtually identical to typical n-type bulk samples. The total thermal conductivity of such nanoparticle pellets is by as much as one order of magnitude smaller than that of the bulk material showing characteristic features of a highly polycrystalline sample. The power factor of 5 $\mu$WK$^{-2}$cm$^{-1}$ is unprecedented by previous samples purely made from solution-grown Bi2Te3 nanoparticles. Optimizing the thermopower by finding suitable grain sizes holds for further improvements.

\section{Experimental}

All manipulations were carried out under an inert atmosphere using standard Schlenck techniques if not stated otherwise. 

\textbf{(I) Synthesis of 7-9 nm bismuth nanoparticles.} In a typical synthesis, bismuth acetate (1.136 g, 3.000 mmol, 99 \% Aldrich) was mixed with 1-dodecanethiol (33.3 mL, 98 \% Aldrich) and heated to 45$^{\circ}$C for 45 min under vacuum on which a transparent, yellow solution was obtained. The flask was flooded with nitrogen, set to ambient pressure and it was heated to 60$^{\circ}$C on which oleylamine (66.7 mL, 70 \%, Aldrich) was quickly added under stirring. The solution turned immediately orange and gradually darkened over the course of several hours. After 24 h the as prepared bismuth nanoparticles were ready for further manipulations.

\textbf{(II) Synthesis of 12-14 nm bismuth nanoparticles.} All manipulations were identical to (I) except for the amount of 1-dodecanethiol (20.0 mL). After the evacuation period and flooding with nitrogen, toluene (40.0 mL, analytical grade, Fluka) was added, followed by oleylamine (40.0 mL). The solution darkened immediately.

\textbf{(III) Synthesis of 40 nm bismuth nanoparticles.} All manipulations were identical to (II) but instead of oleylamine distilled TOP (9.0 mL, 90 \%, Merck) was injected quickly.

\textbf{(IV) Purification of bismuth nanoparticles for characterization.} A fraction of the dark-brown colloidal solution obtained under (I-III) was mixed with ethanol (25 vol-\%, analytical grade, Fluka) and centrifuged at 4500 rpm for 5 min. The light yellow supernatant was removed under nitrogen and the almost black precipitate dissolved in a few drops of chloroform (analytical grade, Fluka). Such a solution can either be purified by one further precipitation cycle with ethanol/chloroform or be treated with a spatula tip of (mPEO2000)2PEI600 ligand. The ligand exchange occurs practically instantaneously on short shaking after which multiple precipitation cycles with hexane as the precipitation agent and chloroform as solvent after centrifugation may be applied. (For more details see supporting information). The purified nanoparticles should be stored in the absence of oxygen to prevent aging. 

\textbf{(V) Preparation of 0.500 M solution of tellurium in TOP (Te@TOP).} In a glovebox, tellurium (1.276 g, 10.00 mmol, 99.999 \%, Chempur) and octadecylphosphonic acid (102 mg, Alfa Aesar) were suspended in distilled TOP (20.0 mL, 90 \%, Merck) under stirring. It was heated stepwise to 220$^{\circ}$C from room temperature by increasing the temperature by approximately 50$^{\circ}$C every hour. The final temperature was kept until a completely transparent, orange solution was obtained which turned to bright-yellow on cooling to room temperature. The solution was stored in the glovebox.

\textbf{(VI) Synthesis of bismuth-tellurium-alloy nanoparticles.} The as synthesized bismuth nanoparticles from (I-III) were used in the same flask without any further purification. For example, (V) (9.0 mL, 4.5 mmol Te) was injected into the product from (I) at 60$^{\circ}$C under stirring. The reaction system was kept under these conditions for 2-3 days. For characterization, the procedures described under (IV) were applied.

\textbf{(VII) Synthesis of Bi2Te3 nanoparticles.} The as synthesized (VI) was heated in the same flask without further purification to 110$^{\circ}$C for 18 h after which it was cooled to room temperature. For characterization the particles were treated according to (IV).

\textbf{(VIII) Ligand removal from Bi2Te3 nanoparticles.} As prepared Bi2Te3 nanoparticles were purified on undergoing three precipitation cycles with ethanol/chloroform as described under (IV) on which the black precipitate was insoluble in chloroform. After removing the supernatant, the precipitate was mixed with a large excess of oleic acid ($\sim$15 mL, 90 \%, Aldrich) and allowed to stir overnight on which a black suspension was formed. The supernatant was removed after short centrifugation and fresh oleic acid was added on which the mixture was allowed to stir for several hours. The supernatant was removed again after centrifugation and it was washed three times with hexane (analytical grade, Aldrich). A few millilitres of fresh hexane were added and the black, insoluble precipitate suspended. An equal volume fraction of distilled hydrazine hydrate (98 \%, Aldrich) was added to form a two-phase system. On stirring overnight followed by centrifugation (4500 rpm, 5 min), practically the entire black precipitate had passed into the hydrazine phase leaving a milky-white precipitate at the bottom of the (upper) hexane phase. If the phase transfer does not occur quantitatively, a few drops of chloroform may be added.

The hexane phase was discarded and the hydrazine hydrate supernatant carefully removed from the black precipitate in the absence of air. It was washed three more times with fresh hydrazine hydrate and hexane until the black precipitate effortlessly went into the hydrazine phase and the hexane phase appeared completely transparent. All solvents were removed and it was dried under vacuum for several hours on which a fine black powder was obtained.
Typically, the starting amounts specified under (I) yield approximately 1 g of Bi2Te3 nanopowder.

\textbf{(IX) Fabrication of Bi2Te3 nanoparticle pellets by spark plasma sintering.} Typically, 100 mg of (VIII) kept under argon were loaded into a WC/Co die of 8.0 mm x 1.5 mm in area. The powder was pressed to a solid pellet of equal dimensions and approximately 1.8 mm in height by spark plasma sintering in a SPS-515 ET/M apparatus (Dr. Sinter lab) under varying conditions. For example, on applying 325 - 358 MPa pressure, the die containing the nanopowder was heated from 20$^{\circ}$C to 50$^{\circ}$C in 5.0 min with no hold time (conditions "c") or 5.0 min hold time (conditions "d") by applying a DC current between 0 - 165 A, and immediately allowed to cool down to room temperature. The obtained Bi2Te3  nanoparticle pellets were mechanically robust and silver-metallic in appearance. 

For the synthesis of Bi2Te3 bulk materials, 100 mg of commercially available Bi2Te3 (99.99 \%) were heated from 20$^{\circ}$C to 300$^{\circ}$C in 30 min with 10 min hold time (conditions "e"). For additional information on temperature/time profiles see supporting information.

\textbf{Characterization.} (HR-)TEM imaging was performed with a Philips CM-300 UT microscope at 200 kV or a JEM-Jeol-1011 microscope at 100 kV with a CCD camera (Gatan 694). SEM images were obtained on a LEO1550 scanning electron microscope with a spatial resolution of $\sim$1 nm.
 
Powder XRDs were recorded using a Philipps XPert-diffractometer with Bragg-Brentano-geometry on applying copper-K$_{\alpha}$ radiation ($\lambda$ = 154.178 pm, U = 45 kV; I = 40 mA).

Dynamic light scattering measurements were carried out with a Malvern Zetasizer Nano-ZS-apparatus equipped with 173$^{\circ}$ back-scattering system and He-Ne-Laser of 633 nm emission wavelength. 

For transport measurements under dc conditions, four 0.25 mm$^{2}$ gold electrodes (40 nm thickness, 1 nm titanium wetting layer below) were vacuum deposited onto a pellet of sintered nanoparticles, with a spacing of 0.5 mm between contact 1 and 2 and 3 and 4, respectively. The spacing between contact 2 and 3 was up to 4.0 mm. A probe station (Lakeshore desert) with micromanipulators and Semiconductor parameter analyzer (Keithley-4200) was applied to measure the transport properties under vacuum (10$^{-3}$ mbar) between 77 K and 300 K.

For transport measurements under ac conditions, four 50 $\mu$m gold wires were glued (conductive silver glue) onto the sample with the geometry specified above in air and allowed to dry for several hours under helium. The gold wires were brazed to a physical property measurement system (Quantum Design) in air and the transport properties measured under helium between 5 K and 300 K. The excitation current was 0.2 mA to 2.0 mA depending on the sample.

For thermopower and thermal conductivity measurements, the sample was equipped with four short copper bars glued to the sample at 150$^{\circ}$C for thirty minutes in an argon atmosphere in a similar geometry as for the transport measurements using a conductive glue (Polytech, EpoTek H20E). The copper bars were contacted to a physical property measurement system (Quantum Design) and the thermoelectric properties measured between 5 K and 300 K by a relaxation-time method and a low-frequency square-wave using two thermometers under a reduced helium atmosphere.

For high temperature measurements of the thermopower and resistivity a ZEM-3 apparatus (ULVAC-RIKO) was applied under a low-pressure helium atmosphere. The thermopower was determined by a static dc method where the resistivity was simultaneously measured by a four-terminal set-up.


\clearpage

\begin{thebibliography}{00}

\bibitem{1} M. H. Ettenberg, W. A. Jesser, F. D. Rosi, in Proc. 15th Int. Conf. on Thermoelectrics (Ed.: T. Caillat), IEEE, Piscataway, NJ 1996, pp. 52–56.
\bibitem{2} L. D. Hicks, M. S. Dresselhaus, Phys. Rev. B 1993, 47, 12727.
\bibitem{3} L. D. Hicks, M. S. Dresselhaus, Phys. Rev. B 1993, 47, 16631.
\bibitem{4} L. D. Hicks, T. C. Harman, X. Sun,M. S. Dresselhaus, Phys. Rev. B 1996, 53, 10493.
\bibitem{5} A. Kitun, A. Balandin, J. L. Liu, K. L. Wang, J. Appl. Phys. 2002, 88, 696.
\bibitem{6} M. S. Dresselhaus, Y.-M. Lin, S. B. Cronin, M. R. Black, O. Rabin, G. Dresselhaus, in Nonlithographic and Lithographic Methods for Nanofabrication: Symposium Proceedings (Ed.: S. Smith), Technomic Publishing Co, Inc., Lancaster, PA 2001.
\bibitem{7} R. Venkatasubramanian, E. Siiwvola, T. Colpitts, B. O'Quinn, Nature 2001, 413, 597.
\bibitem{8} T. C. Harman, P. J. Taylor, M. P. Walsh, B. E. LaForge, Science 2002, 297, 2229.
\bibitem{9} B. Poudel, Q. Hao, Y. Ma, Y. Lan, A. Minnich, B. Yu, X. Yan, D. Wang, A. Muto, D. Vashaee, X. Chen, J. Liu, M. S. Dresselhaus, G. Chen, Z. Ren, Science 2008, 320, 634.
\bibitem{10} N. Gothard, X. Ji, J. He, T. M. Tritt, J. Appl. Phys. 2008, 103, 054314.
\bibitem{11} M. Tokita, J. Soc. Powder Technol. Jpn. 1993, 30, 790.
\bibitem{12} W. Lu, Y. Ding, Y. Chen, Z. L. Wang, J. Fang, J. Am. Chem. Soc. 2005, 127, 10112.
\bibitem{13} A. Purkayastha, S. Kim, D. D. Gandhi, P. G. Ganesan, T. Borca-Tasciuc, G. Ramanath, Adv. Mater. 2006, 18, 2958.
\bibitem{14} A. Purkayastha, Q. Yan, M. S. Raghuveer, D. D. Gandhi, H. Li, Z. W. Liu, R. V. Ramanujan, T. Borca-Tasciuc, G. Ramanath, Adv. Mater. 2008, 20, 2679.
\bibitem{15} M. R. Dirmyer, J. Martin, G. S. Nolas, A. Sen, J. V. Badding, Small 2009, 5, 933.
\bibitem{16} J. W. G. Bos, H. W. Zandbergen, M.-H. Lee, N. P. Ong, R. J. Cava, Phys. Rev. B 2007, 75, 195203.
\bibitem{17} M. V. Kovalenko, M. Scheele, D. V. Talapin, Science 2009, 324, 1417.
\bibitem{18} A. Glatz, I. S. Beloborodov, Phys. Rev. B 2009, 79, 041404(R).
\bibitem{19} D. C. K. MacDonald, E. Mooser, W. B. Pearson, I. M. Templeton, S. B. Woods, Philos. Mag. 1959, 4, 433.
\bibitem{20} F. Yu, J. Zhang, D. Yu, J. He, Z. Liu, B. Xu, Y. Tian, J. Appl. Phys. 2009, 105, 094303.

\end{thebibliography}
\end{document}